# Time-dependent photon generation by superconducting thin-film rings irradiated by coherent microwave fields

A I Agafonov
National Research Centre "Kurchatov Institute", Moscow, 123182 Russia

E-mail: aai@isssph.kiae.ru

**Abstract.** When a thin-film ring with supercurrent is irradiated by coherent microwave field, the field induces coherent oscillations of all the Cooper pairs involved in the supercurrent. The oscillating charged condensate will emit photons with energies determined by the difference of quantized energy levels of the superconducting ring. This effect can be observed only in the rings, the thickness of which is less than both the field skin-depth and the London penetration depth. The probability of the microwave field-induced single-photon and two-photon emission is calculated. The energy - angle distributions of the emitted photons as a function of both the ring sizes and the fluxoid number trapped in the ring, are explored.

## 1. Introduction

The microscopic theory of superconductivity is based on the concept that the superconducting transition is accompanied by spontaneous breaking of the gauge invariance [1-4]. As a result, the superconducting state is characterized by the complex off-diagonal long-range order parameter. Both the amplitude and the phase of the parameter are fixed, but the number of the Cooper pairs is not conserved. The superconducting condensate can be described as the coherent superposition of states $|N>$ with different numbers of the Cooper pairs $N$ [1,2]:

$$\Psi = \sum_N c_N \Psi_N |N>, \qquad (1)$$

where $c_N$'s are real, and $\Psi_N$ is the many-body wave function which describes all the pairs in the same quantum state with the well-defined phase ($\phi$),

$$\Psi_N \propto \exp(iN\phi). \qquad (2)$$

Note that the BCS state can be presented in the form (1)-(2) with the amplitudes $c_N$ peaked up around the average value $\overline{N}$ [5]. The width of this peak $\Delta N (<< \overline{N})$ is the standard deviation of the number of bosons in the condensate or, in other words, the uncertainty in their numbers.

The fact that the superconducting condensate possesses the phase, leads to a number of experimental consequences [5]. One of them is persistent supercurrents that can flow in multiply-connected structures, in particular, in rings in the absence of external magnetic fields. These currents are, in principle, metastable, and can be varied by the quantum jumps, at which the number of the magnetic-flux quanta, trapped in the superconductor, changes by one or several unites. Multiple fluxoid transitions and, respectively, the discrete changes of the supercurrent can be induced by applying an external magnetic field to the ring [6-12]. Also, the quantized decay of the current states can be due to the inelastic magnetic neutron scattering by superconducting rings [13].

When the superconducting thin-film ring is irradiated by a weak low-frequency electromagnetic field, the latter can uniformly penetrate into the ring if the ring thickness is less than the field skin-depth for the superconductor. In addition, if the field is coherent then on the regular circular motion of the Cooper pairs the field imposes coherent oscillations of the condensate as a whole. The oscillating charged condensate will emit photons with the energies determined mainly by the discrete values of the superconducting ring energy. The field-stimulated single-photon generation by the superconducting thin-film rings has been studied in [14]. This effect is, in principal, analogical to the electromagnetic wave emission by an oscillating dipole in classical electrodynamics.

It was shown that under the action of the coherent field the single-photon matrix element of the supercurrent transition is time-dependent. The method used in [14], is based on the expansion of the time-dependent matrix element in the Fourier series. The first term of this series, which does not depend on the time variation of the electromagnetic field, is only taken into account.

It turned out that each channel of the supercurrent decay is characterized by a threshold intensity of the coherent field at which this decay channel becomes allowed. But the intensity of this field which causes the collective transitions of all the Cooper pairs, involved in the supercurrent, varies with time. Therefore, we can expect the time dependence of the supercurrent decay which is accompanied by the emission of photons.

This paper is devoted to investigation of the time evolution of single-photon and two-photon generation by the superconducting thin-film rings irradiated by coherent microwave fields. The field intensity dependence of the supercurrent lifetime is studied. In addition, the energy - angle distributions of the emitted photons as a function of both the ring sizes and the fluxoid number trapped in the ring in the initial state, are explored.

**2. Wave function in supercurrent state**

We consider a flat ring made of a thin superconducting film, with the inner radius $a \gg \lambda$ (where $\lambda$ is the penetration depth) and the outer radius $b$ such that $b - a \gg \lambda$. As justified above, the ring thickness $d$ is small as compared with the penetration depth of the low-frequency field. It is assumed that the temperature is less than $T_c$, and the magnetic field created by the supercurrent in the ring, is small as compared with the first critical field for the type II superconductors.

Properties of a superconducting ring are completely determined by the condensate wave function $\Psi_m$, where $m$ is the number of the magnetic induction flux quanta (fluxoids), trapped in the ring. So, this function is defined the quantization of the magnetic flux $\Phi_m = m\Phi_0$, the discrete values of the total supercurrent $\Phi_m / L$, and the ring energy $E_m = E_0 m^2$ which is composed of the magnetic energy and the condensate kinetic energy. Here $\Phi_0 = \pi\hbar/e$ is the fluxoid, $L$ is the ring self-inductance and $E_0 = \Phi_0^2 / 2L$ is the ring characteristic energy. For the ring considered, the normalized wave function of the superconducting state with $N$ bosons can be written as:

$$\Psi_{Nm}(\boldsymbol{\rho}) = \frac{1}{\sqrt{\Omega_r}} \exp(iN\phi_m), \qquad (3)$$

where $\Omega_r$ is the ring volume, $\phi_m = m\varphi$ is the phase, $\varphi$ is the azimuthal angle in cylindrical coordinates tied to the ring ($a \leq \rho \leq b$ and $-d/2 \leq z \leq d/2$).

Note that the coherent wave packet (1) and (4) is normalized, $\int |\Psi_{Nm}|^2 d\mathbf{r} = 1$, where the integral is taken over the ring volume.

## 3. Phase in quasiclassical approach

$N$ and $\phi_m$ are canonically conjugate variables that implies $\Delta N \Delta \phi_m \approx 1$. In general, $\Delta N$ is large, and, hence, the phase may be treated quasiclassically [1]:

$$\frac{d}{dt}\hbar\nabla\phi_m = 2e\mathbf{F}_0 \cos(\omega_0 t)), \qquad (4)$$

where the right-hand side of (4) is the force on the Cooper pairs, $\mathbf{F}_0$ is the amplitude of the coherent field irradiated the ring, and $\omega_0$ is the field frequency which is assumed to be small as compared with the superconducting gap. Also, we assume that the wavelength of the field is much larger than the outer radius of the ring, so that the field can be considered as uniform. Without loss of generality, the vector $\mathbf{F}_0$ is considered to be directed against the $y$ – axis. Then $\mathbf{i}_\varphi \mathbf{i}_y = \cos\varphi$, where $\mathbf{i}_\varphi$ is the azimuthal unit vector.

From (4) with the initial condition $\phi_m(t=0) = m\varphi$, we find that the phase is:

$$\phi_m(t) = m\varphi - \frac{2eF_0}{\hbar\omega_0}\rho\sin(\varphi)\sin(\omega_0 t). \qquad (5)$$

Of course, this field can change the ring energy in the initial state. To avoid this, we consider the field of low intensities, for which the field correction to the boson velocity $\mathbf{v} = \hbar\nabla\phi_m/m_C$ ($m_C$ is the mass of the Cooper pair) is small. Then, from (6) we have:

$$\xi_0 b/m \ll 1, \qquad (6)$$

where $\xi_0$ is the field parameter,

$$\xi_0 = \frac{2eF_0}{\hbar\omega_0}. \qquad (7)$$

Thus, in the case (6) the potential energy of the condensate in the external field, $U_f = 2eN\mathbf{F}_0\mathbf{r}\cos(\omega_0 t)$, does not influence the ring energy, but only changes the phase (5).

## 4. Transition operator

The Hamiltonian of the system considered, can be expressed as:

$$H = H_0 + V + W + U_f,$$

where $H_0$ the Hamiltonian for the initial supercurrent is state with $N$ bosons and free electromagnetic field, $V$ and $W$ are the interaction operators,

$$V = \int \mathbf{j}\mathbf{A}d\mathbf{r} \qquad (8)$$

and

$$W = \frac{2e^2 N}{m_C}\int \hat{\mathbf{A}}^2 d\mathbf{r}. \qquad (9)$$

Here $\hat{\mathbf{j}}$ is the operator of the superconducting current density in the ring, $\mathbf{A}$ is the vector-potential of the electromagnetic field generated by the superconducting ring, $m_C$ and $2e$ are the mass and charge of the Cooper pair.

As discussed in section 3, at low microwave intensities satisfied (6), one can neglect the field correction to the ring energy. It means that the commutator of the potential energy of the Cooper pairs in the classical electromagnetic field $U_f$ and the Hamiltonian $H_0$ is small. Then the evolution operator satisfies the equation:

$$i\hbar\frac{\partial}{\partial t}S(t,0) = \left[\exp\left(\frac{i}{\hbar}H_0 t\right)(V+W)\exp\left(-\frac{i}{\hbar}H_0 t\right) + U_f\right]S(t,0) \qquad (10)$$

with $S(0,0) = 1$.

The supercurrent decay in the ring with the single-photon emission is determined by the matrix element of (8) in the first-order perturbation theory. Two-photon decay of the supercurrent is given by the composite matrix element of the operator (8) in the second order, describing the process of the sequential emission of two photons, and the matrix element of (9) in the first order, which represents the simultaneous emission of two photons. Below we justify the claim that the operator (9) defines, in general, the two-photon emission for the macroscopic ring.

Using (10), in the first order over the interactions (8) and (9) we obtain:

$$S_1(t,0) = -\frac{i}{\hbar}\int_0^t dt_1 \exp\left(\frac{i}{\hbar}\left[H_0 t_1 - \int_{t_1}^t U_f(t_2)dt_2\right]\right)(V+W)\exp\left(-\frac{i}{\hbar}\left[H_0 t_1 + \int_0^{t_1} U_f(t_2)dt_2\right]\right). \quad (11)$$

The term with the operator $W$ in the right-hand side of (11) can be simplified because (9) commutes with the classical field $U_f$. The interaction (8) does not commute with $U_f$. However, the field correction to the supercurrent, which is due to the oscillation of particles in the low-frequency field, is small for the intensities satisfied (6), as shown below.

It is easy to see that the action of $\exp\left(-\frac{i}{\hbar}\int_0^t U_f(t_1)dt_1\right)$ on the wave function (3) results in the change of the phase that corresponds to (5).

## 5. Single-photon emission

Using (3) and (11), the amplitude of the supercurrent transition $m \to m_1$ with the single-photon emission is:

$$<\mathbf{k},\Psi_{Nm_1}|S_1|\Psi_{Nm},0> = -\frac{i}{\hbar}\int_0^t V_{m_1,m}^{\mathbf{k},0}(t,t_1)\exp\left(-\frac{it_1}{\hbar}(E_m - E_{m_1} - \hbar\omega_k)\right)dt_1 . \quad (12)$$

In the right-hand side of (12) the matrix element of (8) is written as:

$$V_{m_1,m}^{\mathbf{k},0} = \left(\frac{\hbar}{2\varepsilon_0\omega_k}\right)^{1/2}\int \mathbf{l}_{\mathbf{k}\sigma}\exp(-i\mathbf{kr})\left(\Psi_{Nm_1}^*\exp\left[-\frac{i}{\hbar}\int_{t_1}^t dt' U_f(t')\right]\hat{\mathbf{j}}\exp\left[-\frac{i}{\hbar}\int_0^{t_1} dt' U_f(t')\right]\Psi_{Nm}\right)d\mathbf{r}, \quad (13)$$

where $\mathbf{l}_{\mathbf{k}\sigma}$, $\mathbf{k}$ and $\hbar\omega_k$ are the photon polarization, wave vector and energy, respectively.

Using the well-known expression for the current density operator [5,14], the single-photon matrix element (13) is formed as:

$$V_{m_1,m}^{\mathbf{k},0} = \tilde{V}_{m_1,m}(\mathbf{k})(I_1(t,N) - I_2(t,t_1,N)) \quad (14)$$

with

$$\tilde{V}_{m_1 m}(\mathbf{k}) = \frac{2^{1/2}e}{m_C \Omega_r}*\left(\frac{\hbar^3}{\varepsilon_0\omega_k}\right)^{1/2}*\frac{\sin(\frac{kd}{2}\cos\theta_k)}{k\cos\theta_k}, \quad (15)$$

where $\theta_k$ is the polar angle of the photon wave vector. Taking advantage of $\mathbf{i}_\varphi \mathbf{l}_{\mathbf{k}\sigma} = \sin(\theta_l)\sin(\varphi_l - \varphi)$ and $\mathbf{i}_y \mathbf{l}_{\mathbf{k}\sigma} = \sin(\theta_l)\sin(\varphi_l)$, where $\theta_l$ and $\varphi_l$ are the polar and azimuthal angles of the photon polarization $\mathbf{l}_{\mathbf{k}\sigma}$, Eq. (14) are introduced the notations:

$$I_1(t,N) = N(m+m_1)\sin(\theta_l)\int_a^b d\rho \int_0^{2\pi} d\varphi \sin(\varphi_l - \varphi)$$
$$\exp\left(-ik_\rho\rho\cos(\varphi - \varphi_k) + iN[(m-m_1)\varphi - \xi_0\rho\sin(\varphi)\sin(\omega_0 t)]\right), \quad (16)$$

$$I_2 = N\xi_0[2\sin(\omega_0 t_1) - \sin(\omega_0 t)]\sin(\theta_l)\sin(\varphi_l)\int_a^b \rho d\rho \int_0^{2\pi} d\varphi$$
$$\exp\left(-ik_\rho\rho\cos(\varphi - \varphi_k) + iN[(m-m_1)\varphi - \xi_0\rho\sin(\varphi)\sin(\omega_0 t)]\right). \quad (17)$$

Here $k_\rho = k\sin(\theta_k)$ and $\varphi_k$ is the azimuthal angle of the photon wave vector.
The right-hand side of (16) is reduced to:

$$I_1 = -N(m+m_1)e^{-iN(m-m_1)\varphi_g}\sin(\theta_l)\int_a^b d\rho \int_0^{2\pi} d\varphi \sin(\varphi - \varphi_l - \varphi_g)e^{iN(m-m_1)\varphi - iN\xi_0 g(t)\rho \sin\varphi}, \quad (18)$$

where

$$g(t) = \left(\sin^2(\omega_0 t) + 2\frac{k_\rho \sin\varphi_k}{N\xi_0}\sin(\omega_o t) + \left(\frac{k_\rho}{N\xi_0}\right)^2\right)^{1/2}. \quad (19)$$

and

$$\sin\varphi_g = \frac{k_\rho \cos\varphi_k}{N\xi_0 g(t)}.$$

After the integration over the azimuthal angle in (18), we obtain:

$$I_1(t,N) = i\pi N(m+m_1)e^{-iN(m-m_1)\varphi_g}\sin(\theta_l)\int_a^b d\rho$$
$$\left(e^{-i(\varphi_l+\varphi_g)}J_{N(m-m_1)+1}(N\xi_0 g(t)\rho) - e^{i(\varphi_l+\varphi_g)}J_{N(m-m_1)-1}(N\xi_0 g(t)\rho)\right), \quad (20)$$

where $J_{N(m-m_1)\pm 1}(N\xi_0 g(t)\rho)$ is the Bessel function with very large order $N(m-m_1)\pm 1$. This function does not vanish only when its argument is comparable with the order.
In (20) we pass to a new variable of integration $z$ defined by

$$z = 2^{1/3}(N(m-m_1)\pm 1)^{2/3}\left(\frac{\xi_0 g(t)\rho}{m-m_1 \pm N^{-1}} - 1\right),$$

and use the asymptotic expansion the Bessel functions with large orders in terms of the Airy functions [15]:

$$J_{N(m-m_1)\pm 1}(N\xi_0 g(t)\rho) = 2^{1/3}(N(m-m_1)\pm 1)^{-1/3}Ai(-z) + O\left((N(m-m_1))^{-1}\right).$$

Taking into account that the number of Cooper pairs $N >>> 1$ in the superconductor, we obtain:

$$I_1(t) = 2\pi \frac{(m+m_1)}{\xi_0 g(t)}e^{-iN(m-m_1)\varphi_g}\sin(\theta_l)\sin(\varphi_l+\varphi_g)\int_{A(t)}^{B(t)}Ai(-z)dz, \quad (21)$$

where the time-dependent limits of integration ($A < B$), are

$$A(t) = 2^{1/3}(N(m-m_1))^{2/3}\left[\frac{\xi_0 a g(t)}{m-m_1} - 1\right] \quad (22)$$

and

$$B(t) = 2^{1/3}(N(m-m_1))^{2/3}\left[\frac{\xi_0 b g(t)}{m-m_1} - 1\right]. \quad (23)$$

The similar calculation approach to (17) gives us:

$$I_2(t) = 2\pi \frac{m-m_1}{\xi_0 g^2(t)}[2\sin(\omega_0 t_1) - \sin(\omega_0 t)]e^{-iN(m-m_1)\varphi_g}\sin(\theta_l)\sin(\varphi_l)\int_{A(t)}^{B(t)}Ai(-z)dz. \quad (24)$$

Because the modulus of the matrix element (14) is of interest, the common phase factor $e^{-iN(m-m_1)\varphi_g}$ in (21) and (24) can be omitted.
The limits of integration (22) and (23) can be large in absolute value, and vary sharply with time. At non-positive values of $z$ the Airy function $Ai(-z)$ decreases exponentially with $|z|$. In this case the integration over $z$ in (21) and (24) can be restricted to the interval $[-\eta_1, 0]$ where $\eta_1 \approx 7$, and $\int_{-\eta_1}^{0}Ai(-z) = \frac{1}{3}$. For the positive values $z$ the Airy function $Ai(-z)$ oscillates with $z$, and the oscillation amplitude decreases. Here the integration over $z$ in (21) and (24) can be restricted to the interval $[0, \eta_2 \approx 10^2]$, where $\int_0^{\eta_2} Ai(-z)dz = \frac{2}{3}$. Because $N >>> 1$, the time-dependent limits $A(t)$ and $B(t)$ will quickly intersect the interval $[-\eta_1, \eta_2]$ with time variation.

To estimate the intersection time, we consider the $g(t)$ function (19) which must be non-zero, otherwise both the lower and upper limits of integration are large and negative, and, hence, $I_{1,2} = 0$. Let the flat ring has the outer radius $b = 2\mu m$, inner radius $a = 1\mu m$ and thickness $d = 800 A^0$. With the density of Cooper pairs $10^{21} cm^{-3}$ in the superconductor, we obtain the number of Cooper pairs $\overline{N} = 10^{21} \pi d(b^2 - a^2) = 0.75*10^9$. According to (23), the upper limit of integration can be positive in certain time intervals if the value $\xi_0 > (m - m_1)/b$, where $m - m_1 \geq 1$ is the change of the fluxoid number in the initial and final states. Let $\xi_0 = 2b^{-1}$. When changing the ring energy $E_m - E_{m_1} = 200 eV$, we have $k_\rho \leq 10^7 cm^{-1}$ and the ratio:

$$\frac{k_\rho}{N\xi_0} \leq 1.3*10^{-6}.$$

Therefore, we can neglect the terms containing $k_\rho$ in the right-hand side of (19). As a result, we get:

$$g(t) = |\sin(\omega_0 t)|. \qquad (25)$$

The upper limit of integration (23) is positive when

$$g(t) > \frac{m - m_1}{\xi_0 b}.$$

Taking into account this condition, we have

$$|\sin\varphi_g| = \frac{k_\rho |\cos\varphi_k|}{N\xi_0 g(t)} < \frac{k_\rho b}{N(m - m_1)}.$$

For the ring considered, $|\sin\varphi_g| < 2.7*10^{-6}$, so that $\varphi_g = 0$ or $\pi$. Hence, the angle $\varphi_g$ in (21) can be omitted.

From (25) we can estimate the time at which the limits of integrations $A(t)$ and $B(t)$ pass through the interval $[-\eta_1, \eta_2]$:

$$\omega_0 \Delta t \propto \frac{\eta_1 + \eta_2}{N^{2/3}} << 1.$$

Neglecting transition regions with such small times and using (25), the expressions (21) and (24) are reduced to the forms:

$$I_1(t) = 2\pi \frac{m + m_1}{\xi_0} \sin(\theta_l)\sin(\varphi_l)s_1(t), \qquad (26)$$

$$I_2(t) = 2\pi \frac{m - m_1}{\xi_0} [2\sin(\omega_0 t_1) - \sin(\omega_0 t)]\sin(\theta_l)\sin(\varphi_l)s_2(t), \qquad (27)$$

where

$$s_{n=1,2}(t) = |\sin\omega_0 t|^{-n} \theta(|\sin\omega_0 t| - \chi_b)\theta(\chi_a - |\sin\omega_0 t|). \qquad (28)$$

Here $\theta$ is the Heaviside step function, $\chi_b = \frac{m - m_1}{\xi_0 b}$ and $\chi_a = \frac{m - m_1}{\xi_0 a}$. In the weak fields given by (6), we have $I_2 << I_1$. Therefore we can limit ourselves the consideration of $I_1$ (26) only.

Note that the function $I_1(t)$ given by (26), does not depend on the number of Cooper pairs in the superconductor that consists with the result of [14]. Considering (16), we can conclude that the perturbation of the superconducting condensate by the coherent low-frequency electromagnetic field can effectively compensate the function $\exp(iN(m-m_1)\varphi)$ with very large global phase. It leads to finite, really measurable lifetimes of the current states in superconducting thin-film rings of relatively small sizes, as shown below.

Using (12)-(14) and (26), the probability of the supercurrent decay in the ring for the transition channel $m \to m_1$ with the single-photon emission is written as:

$$w^s_{m_1 m}(t) = \frac{(2\pi)^3}{\hbar} \frac{(m+m_1)^2}{\xi_0^2} s_1^2(t) \sum_N c_N^2 \sum_{\mathbf{k}} |\widetilde{V}_{m_1 m}(\mathbf{k})|^2 < \sin^2(\theta_l) \sin^2(\varphi_l) >_{pol} \delta(E_m - E_{m_1} - \hbar\omega_k), \quad (29)$$

where $<...>_{pol} = \pi^{-1} \int_0^{2\pi} d\varphi_l ...$ means the average over the photon polarization.

Using $\mathbf{k}\mathbf{l}_{\mathbf{k}\sigma} = 0$, and the real polarization vectors, we have:

$$\sin^2(\theta_l) = \left(1 + tg^2(\theta_k)\cos^2(\varphi_l - \varphi_k)\right)^{-1}.$$

Hence, the average over the photon polarization is reduced to calculation of the function:

$$D_1(\theta_k, \varphi_k) = <\sin^2(\theta_l)\sin^2(\varphi_l)>_{pol} = \pi^{-1} \int_0^{2\pi} d\varphi_l \frac{\sin^2(\varphi_l)}{1 + tg^2(\theta_k)\cos^2(\varphi_l - \varphi_k)}.$$

Integrating over $\varphi_l$, we obtain:

$$D_1 = (1 + \cos 2\varphi_k)|\cos\theta_k| - \frac{2\cos(2\varphi_k)}{tg^2\theta_k}(1 - |\cos\theta_k|). \quad (30)$$

To sum over $N$ in (29), we note that the energy of the ring and, respectively, the characteristic energy $E_0$ are proportional to the Cooper pair number which is not fixed in the superconductor. Since $E_0 \propto R_L^{-1}$, where $R_L = L/\mu_0$ is the ring inductance in unit of $\mu_0$, we can introduce $R_L(N) = \overline{R}_L \overline{N}/N$, where $\overline{R}_L$ corresponds to the average number of Cooper pairs. Also, the energy of emitted photon depends on $N$ and, according to (29), is equal to $\hbar\omega_k(N) = \frac{N}{\overline{N}} E_0(\overline{N})(m^2 - m_1^2)$. Considering the Cooper pair number as a fluctuating quantity, the emitted photon energy is fluctuating as well.

Using (30) and integrating over the azimuthal angle $\varphi_k$, the probability (29) is rewritten as:

$$w^s_{m_1 m}(t) = \frac{2^5 \alpha^2}{\pi} \left(\frac{2m_e}{m_C}\right)^2 s_1^2(t) \frac{m+m_1}{m-m_1} \frac{\lambdabar_e^2 c \overline{R}_L}{d^2(b^2-a^2)^2 \xi_0^2} F_1(\beta), \quad (31)$$

where

$$F_1(\beta) = \sum_N \frac{\overline{N}}{N} c_N^2 \int_0^\pi \frac{\sin\theta_k d\theta_k}{|\cos\theta_k|} \sin^2\left(\frac{N}{\overline{N}} \beta \cos\theta_k\right). \quad (32)$$

Here $\alpha$ is the fine structure constant, $m_e$ is the electron mass, $\lambdabar_e$ is the Compton wavelength of the electron, $c$ is the speed of light, $E_d = 2c\hbar/d$ and

$$\beta = \frac{E_0(\overline{N})(m^2 - m_1^2)}{E_d}. \quad (33)$$

Due to fluctuations of the number of Cooper pairs in the superconducting state, the energy distribution of photons emitted by a certain ring for the same supercurrent decay channel $m \to m_1$, will be a narrow peak determined mainly by the coefficients $c_N$. Passing from summation over $N$ to integration over the photon energy, from (32) we obtain $F_1 = \int P(\hbar\omega_k) d\hbar\omega_k$, where $P(\hbar\omega_k)$ is defined by statistics of the emitted photon energy:

$$P(\hbar\omega_k) = \frac{\overline{N}}{\hbar\omega_k} c^2\left(N = \frac{\overline{N}\hbar\omega_k}{\beta E_d}\right)\left[\gamma + \ln\left(\frac{2\hbar\omega_k}{E_d}\right) - Ci\left(\frac{2\hbar\omega_k}{E_d}\right)\right],$$

where $\gamma$ is the Euler constant, $c_N^2$ is the probability that the number of Cooper pairs is equal to $N$ in the superconducting ring.

Note that there is a principal possibility to determine the coefficients $c_N$ by carrying out statistical measurements of the energies of photons emitted by the same ring. Detection of this photon peak may provide a new approach for the study of fluctuation phenomena in superconductors.

In [1-2] the wave packet (1) was formed with $c_N$'s given by the normal Gaussian distribution. In this case, (32) is reduced to:

$$F_1(\beta) = \int_0^\pi \frac{\sin\theta_k d\theta_k}{|\cos\theta_k|} \sin^2(\beta\cos\theta_k) + O\left((\Delta N/\overline{N})^2\right). \tag{34}$$

Because the ratio $\Delta N/\overline{N} \ll 1$, we can neglect the correction to the term in (34) with the average number of Cooper pairs.

Note that the function under the integral sign in the right-hand side of (34) determines the polar angle distribution of photons.

The probability (31) is time dependent. We average the probability (31) over the field oscillation period:

$$<w_{m_1 m}^s> = \frac{\omega_0}{2\pi} \int_0^{2\pi/\omega_0} w_{m_1 m}^s(t)dt.$$

Using (28), we obtain:

$$<w_{m_1 m}^s> = \frac{2^6 \alpha^2}{\pi^2}\left(\frac{2m_e}{m_C}\right)^2 \frac{\lambdabar_e^2 c \overline{R}_L}{d^2 b^3(1-(a/b)^2)^2} \frac{m+m_1}{(m-m_1)^2 \xi_0} F_1(\beta) q_1(\chi_a, \chi_b), \tag{35}$$

where

$$q_1(\chi_a, \chi_b) = \begin{cases} (1-\chi_b^2)^{1/2}, & \chi_b < 1 \text{ and } \chi_a > 1 \\ (1-\chi_b^2)^{1/2} - \frac{a}{b}(1-\chi_a^2)^{1/2}, & \chi_a < 1 \end{cases}.$$

The averaged probability (35) defines the lifetime of the supercurrent in its decay channel $m \to m_1$ with the single-photon emission:

$$\tau_{m_1 m}^s = <w_{m_1 m}^s>^{-1}.$$

**6. Two-photon emission**

The single-photon matrix element (15) of the interaction operator (8) is inversely proportional to the ring volume, $\Omega_r$. Hence, the composite matrix element of (8) which describes the sequential emission of two photons, $\propto \Omega_r^{-2}$. The matrix element of the interaction (9) that represents the simultaneous emission of two photons, $\propto \Omega_r^{-1}$, as we see below. Thus, for macroscopic rings the two-photon generation will be determined by the interaction (9).

Taking into account (3) and (11), the amplitude of the supercurrent transition $m \to m_1$ with the two-photon emission is:

$$<\mathbf{k}_1\mathbf{k}_2, \Psi_{Nm_1}|S_1|\Psi_{Nm}, 0> = -\frac{i}{\hbar}\int_0^t W_{m_1,m}^{\mathbf{k}_1\mathbf{k}_2,0} \exp\left(-\frac{it_1}{\hbar}(E_m - E_{m_1} - \hbar\omega_{k_1} - \hbar\omega_{k_2})\right)dt_1. \tag{36}$$

In (36) the matrix element of the interaction (9) can be written as:

$$W_{m_1,m}^{\mathbf{k}_1\mathbf{k}_2,0}(N) = \widetilde{W}_{m_1,m}(\mathbf{k}_1,\mathbf{k}_2) * I_3(N,t), \tag{37}$$

where

$$\widetilde{W}_{m_1,m}(\mathbf{k}_1,\mathbf{k}_2) = \frac{2e^2}{m_C \Omega_r} * \frac{\hbar(\mathbf{l}_{\mathbf{k}_1\sigma_1}\mathbf{l}_{\mathbf{k}_2\sigma_2})}{\varepsilon_0 \sqrt{\omega_{k_1}\omega_{k_2}}} \frac{\sin(\frac{d}{2}k_z)}{k_z}, \tag{38}$$

and

$$I_3(N,t) = N \int_a^b \rho d\rho \int_0^{2\pi} d\varphi \, \exp\bigl(iN(m-m_1)\varphi - iN\xi_0 \rho \sin(\varphi)\sin(\omega_0 t) - ik_\rho \rho \cos(\varphi - \varphi_k)\bigr). \tag{39}$$

Here $\mathbf{k} = \mathbf{k}_1 + \mathbf{k}_2$ is the total wave vector of the photons, $k_z = k\cos\theta_k = k_1\cos\theta_{k_1} + k_2\cos\theta_{k_2}$ and $k_\rho = k\sin\theta_k$ are the wave vector components, $\mathbf{l}_{\mathbf{k}_1\sigma_1}$ and $\mathbf{l}_{\mathbf{k}_2\sigma_2}$ are the polarization vectors of the photons with the energies $\hbar\omega_{k_1}$ and $\hbar\omega_{k_2}$, $\theta_{k_1}, \theta_{k_2}$ and $\varphi_{k_1}, \varphi_{k_2}$ are the polar and azimuthal angles of the photon wave vectors $\mathbf{k}_1$ and $\mathbf{k}_2$.

Except for the factor $\xi_0[2\sin(\omega_0 t_1) - \sin(\omega_0 t)]\sin(\theta_l)\sin(\varphi_l)$ before the integral sign in (17), $I_3$ (38) and $I_2$ have the same form. For this reason, we immediately obtain:

$$I_3(t) = 2\pi \frac{m-m_1}{\xi_0^2} s_2(t), \tag{40}$$

where the time dependent function $s_2(t)$ is given by (28).

Using (36)-(38) and (40), the probability of the supercurrent decay in the ring for the transition channel $m \to m_1$ with the simultaneous emission of two photons is written as:

$$w_{m_1 m}^t(t) = \frac{(2\pi)^3}{\hbar} \frac{(m-m_1)^2}{\xi_0^4} s_2^2(t) \sum_N c_N^2 \sum_{\mathbf{k}_1,\mathbf{k}_2} <|\widetilde{W}_{m_1,m}(\mathbf{k}_1,\mathbf{k}_2)|^2>_{pol} \delta(E_m - E_{m_1} - \hbar\omega_{k_1} - \hbar\omega_{k_2}), \tag{41}$$

where $<...>_{pol} = (2\pi)^{-2} \sum_{\sigma_1,\sigma_2} \int_0^{2\pi} d\varphi_{l_1} \int_0^{2\pi} d\varphi_{l_2} ...$ means the average over the photon polarizations, $\varphi_{l_1}$ and $\varphi_{l_2}$ are the azimuthal angles of the polarization vectors of the photons.

Taking into account (38), this average is reduced to calculation of the two-photon correlation function:

$$D_2(\theta_{k_1},\varphi_{k_1},\theta_{k_2},\varphi_{k_2}) = (2\pi)^{-2} \sum_{\sigma_1,\sigma_2} \int_0^{2\pi} d\varphi_{l_1} \int_0^{2\pi} d\varphi_{l_2} (\mathbf{l}_{\mathbf{k}_1\sigma_1}\mathbf{l}_{\mathbf{k}_2\sigma_2})^2. \tag{42}$$

Using $\mathbf{k}_i \mathbf{l}_{\mathbf{k}_i \sigma_i} = 0$, $i = 1,2$, and the real polarization vectors, we obtain:

$$\mathbf{l}_{\mathbf{k}_1\sigma_1} \mathbf{l}_{\mathbf{k}_2\sigma_2} = \frac{tg(\theta_{k_1})\cos(\varphi_{k_1} - \varphi_{l_1}) tg(\theta_{k_2})\cos(\varphi_{k_2} - \varphi_{l_2}) + \cos(\varphi_{l_1} - \varphi_{l_2})}{\sqrt{1 + tg^2(\theta_{k_1})\cos^2(\varphi_{k_1} - \varphi_{l_1})} \sqrt{1 + tg^2(\theta_{k_2})\cos^2(\varphi_{k_2} - \varphi_{l_2})}}. \tag{43}$$

Substituting (43) in (42), the integrations over the angles $\varphi_{l_1}$ и $\varphi_{l_2}$ result in:

$$D_2 = 2|\cos\theta_{k_1}\cos\theta_{k_2}| + 4(1-|\cos\theta_{k_1}|)(1-|\cos\theta_{k_2}|)\left(1 + \frac{2\cos(\varphi_{k_1}-\varphi_{k_2})}{tg\theta_{k_1} tg\theta_{k_2}}\right) +$$

$$8\cos 2(\varphi_{k_1} - \varphi_{k_2}) \frac{1-(1+\frac{1}{2}tg^2\theta_{k_1})|\cos\theta_{k_1}|}{tg^2\theta_{k_1}} * \frac{1-(1+\frac{1}{2}tg^2\theta_{k_2})|\cos\theta_{k_2}|}{tg^2\theta_{k_2}}. \tag{44}$$

It is not hard to make sure that the sum over $N$ in (41) is reduced to:

$$\sum_N \frac{N}{\overline{N}} c_N^2 \sin^2\left(\beta \frac{N}{\overline{N}}(\gamma(\cos\theta_{k_1} - \cos\theta_{k_2}) + \cos\theta_{k_2})\right), \tag{45}$$

where $\beta$ is given be (33), and $\gamma = \hbar\omega_{k_1}/E_0(\overline{N})(m^2 - m_1^2)$ is the ratio of the emitted photon energy to the ring energy change due to the transition $m \to m_1$.

If $c_N$'s are given by the normal Gaussian distribution then the sum (45) results in:

$$\sin^2(\beta(\gamma(\cos\theta_{k_1} - \cos\theta_{k_2}) + \cos\theta_{k_2})) + O((\Delta N/\overline{N})^2).$$

Since $\Delta N/\overline{N} \ll 1$, we can omit the corrections to the term with the average number of Cooper pairs.

Taking into account (44), the integrations in (41) over the azimuthal angles $\varphi_{k_1}$ and $\varphi_{k_2}$ are easily carried out. Thereafter the supercurrent decay probability (41) is reduced to the form:

$$w_{m_1 m}^t(t) = 2\alpha \left(\frac{2m_e}{m_C}\right)^2 s_2^2(t) \frac{\lambda_e^2 c (m+m_1)(m-m_1)^3}{\overline{R}_L d^2 (b^2 - a^2)^2 \xi_0^4} F_2(\beta) \tag{46}$$

with

$$F_2(\beta) = \int_0^1 \gamma(1-\gamma) d\gamma \int_0^\pi \sin\theta_{k_1} d\theta_{k_1} \int_0^\pi \sin\theta_{k_2} d\theta_{k_2} \frac{\sin^2(\beta(\gamma(\cos\theta_{k_1} - \cos\theta_{k_2}) + \cos\theta_{k_2}))}{(\gamma(\cos\theta_{k_1} - \cos\theta_{k_2}) + \cos\theta_{k_2})^2}$$

$$(3|\cos\theta_{k_1}\cos\theta_{k_2}| + 2(1 - |\cos\theta_{k_1}| - |\cos\theta_{k_2}|)). \tag{47}$$

Note that the function under the integrals in the right-hand side of (47) determines both the energy and polar angle distributions of the emitted photons.

It is convenient to average the time-dependent probability (46) over the field oscillation period. Using (28), we obtain:

$$< w_{m_1 m}^t > = \frac{4\alpha}{3\pi} \left(\frac{2m_e}{m_C}\right)^2 \frac{\lambda_e^2 c}{\overline{R}_L d^2 b (1 - (a/b)^2)^2} \frac{(m+m_1)}{\xi_0} F_2(\beta) q_2(\chi_a, \chi_b), \tag{48}$$

where

$$q_2 = \begin{cases} (1 + 2\chi_b^2)(1 - \chi_b^2)^{1/2}, & \chi_b < 1 \text{ and } \chi_a > 1 \\ (1 + 2\chi_b^2)(1 - \chi_b^2)^{1/2} - \frac{a^3}{b^3}(1 + 2\chi_a^2)(1 - \chi_a^2)^{1/2}, & \chi_a < 1 \end{cases}.$$

The averaged probability (48) allows us to introduce the lifetime of the supercurrent in its decay channel $m \to m_1$ with the two-photon emission:

$$\tau_{m_1 m}^t = < w_{m_1 m}^t >^{-1}.$$

### 7. Numerical results and discussion

First of all note a surprising feature that the probability (31) of the single-photon decay of the supercurrent is proportional to the square of the fine structure constant ($\alpha$), while the two-photon probability (46) is proportional to $\alpha$. This is due to the fact that the "quantum" energy of superconducting rings is inversely proportional to the square of the electron charge, $E_0 \propto e^{-2}$.

Given the time-dependent functions $s_n(t)$ (28), we conclude that both the single-photon and two-photon generation by the superconducting rings has a threshold dependence on the coherent field parameter $\xi_0$ (7). The generation threshold is defined by $\xi_0^{th} = b^{-1}$. At $\xi_0 < \xi_0^{th}$ all the supercurrent transitions $m \to m_1$ are forbidden. It means that the supercurrent in the ring is persistent, and photon emissions do not occur.

The field parameter $\xi_0$ increases with increasing the coherent field intensity or decreasing the field frequency. When $\xi_0^{th} < \xi_0 < 2\xi_0^{th}$, the only supercurrent transition $m \to m-1$ is allowed with destruction of one fluxoid in the final state of the superconducting ring. For $m$ fluxoids in the initial state, the ring energy change is equal to $E_0(2m-1)$. This energy can be carried away by either a single photon with the probability (31) or two photons with the probability (46). Thus, the single-photon and two-photon channels of the supercurrent decay are characterized by the same threshold.

At $2\xi_0^{th} < \xi_0 < 3\xi_0^{th}$ besides the already allowed transition $m \to m-1$, the next supercurrent transition $m \to m-2$ becomes allowed. When $n\xi_0^{th} < \xi_0 < (n+1)\xi_0^{th}$, the number of the allowed supercurrent transitions is equal to $n$ with destruction of 1, 2, 3,... $n$ fluxoids in the final state. Of course, the probabilities of these transitions vary with the field parameter.

Calculations of the probabilities of supercurrent decay with the single-photon and two-photon emission are carried out for two flat thin-film rings I and II, parameters of which are presented in table 1. The thickness of these rings is assumed to be less than the field skin-depth for the superconductor. We use the results for the self-inductance $L/\mu_0 a$ of the superconducting thin-film rings as a function of the ratio $a/b$, shown in Fig. 2 of Ref. [16]. The parameter of the curves presented, is $\Lambda/b$, where $\Lambda = \lambda^2/d$ is the two-dimensional effective penetration depth, $\lambda$ is the London penetration depth, and the ring thickness $d < \lambda/2$. The penetration depth is taken $\lambda = 2.2*10^3 A^0$ that is a typical value for type-II superconductors. Then, from the given curves we calculate the self-inductance coefficient $\kappa = \overline{R}_L/a$ and the characteristic energy $E_0$ of these rings. Hereafter $m_C = 2m_e$ is used.

Table 1. The self-inductance coefficient $\kappa = \overline{R}_L/a$ and the characteristic energy $E_0$ for two thin-film rings with the hole radius $a$ and outer radius $b$. The thickness of the rings $d = 800 A^0$.

| N | $b(\mu m)$ | $a(\mu m)$ | $\kappa$ | $E_0 (eV)$ |
|---|---|---|---|---|
| I | 2 | 1 | 8.29 | 1.28 |
| II | 20 | 10 | 3.25 | 0.33 |

Let us estimate the threshold intensity of the coherent microwave field which corresponds to $\xi_0^{th} = b^{-1}$. Using (7), we have

$$I_{th} = \frac{c\varepsilon_0 \hbar^2 \omega_0^2}{8e^2 b^2}.$$

For the ring I with the outer radius $b = 2\mu m$ and the field frequency $\omega_0 = 10^{12} s^{-1}$ we obtain $I_{th} = 3.56 mW/cm^2$. The threshold intensity decreases with increasing the ring radius and with decreasing the coherent field frequency.

According to (28), during the coherent field action the single-photon and two-photon emission takes place only at certain times, when

$$\frac{m - m_1}{\xi_0 b} < |\sin \omega_0 t| < \frac{m - m_1}{\xi_0 a}.$$

Thus, the photon generation by the superconducting ring has a time-pulsing character. Figure 1 demonstrates the time dependence of the two-photon generation at various values of the field parameter. When the field parameter is slightly higher than the generation threshold value $\xi_0^{th}$, there are narrow temporal regions near the values $\omega_0 t = \pi/2$ and $\omega_0 t = 3\pi/2$ where the photon emission occurs (curve 1 in figure 1). With increasing the field parameter these regions expand, as it shown by the curves 2 and 3 in figure 1. This leads to more rapid decay of the supercurrent in the ring with the two-photon emission. When this parameter exceeds the value $(b/a)\xi_0^{th}$, dips of the photon generation curves are appeared near the values $\omega_0 t = \pi/2$ and $\omega_0 t = 3\pi/2$ (curve 4 in figure 1). Then, these dips increase with $\xi_0$, and the temporal regions of the photon generation are narrowed (curve 5 in figure 1). Accordingly, decay of the supercurrent becomes less probable. When the field parameter $\xi_0 > 2\xi_0^{th}$, the next supercurrent transition becomes allowed (dashed curves 6 and 7 in figure 1).

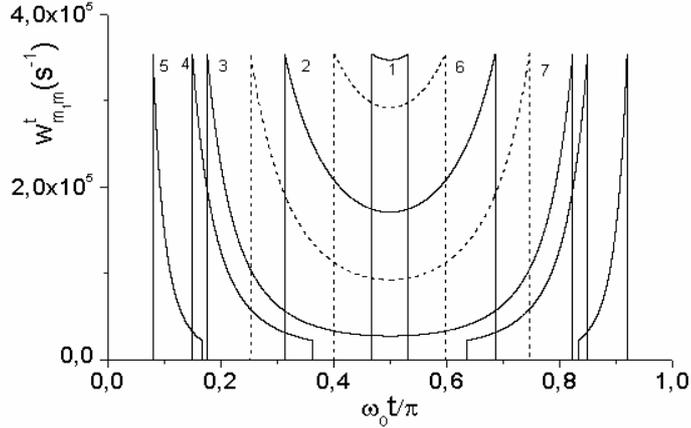

Figure 1. The time dependence of the two-photon emission from the ring I. The fluxoid number in the initial superconducting state is equal to 101. The solid curves correspond to destruction of one fluxoid in the final state: curve 1 - $\xi_0 = 1.005\xi_0^{th}$; curve 2 - $\xi_0 = 1.2\xi_0^{th}$; curve 3 - $\xi_0 = 1.9\xi_0^{th}$; curve 4 - $\xi_0 = 2.2\xi_0^{th}$; curve 5 - $\xi_0 = 4\xi_0^{th}$. The dash curves correspond to destruction of two fluxoids in the final state: curve 6 - $\xi_0 = 2.1\xi_0^{th}$ and curve 7 - $\xi_0 = 2.8\xi_0^{th}$

Note that the field parameter dependence of the temporal dynamics of the single-photon generation is the same as discussed above.

Figure 2 shows the field parameter dependence of the averaged probabilities of the single photon generation by the ring I. The curve 1 corresponds to the emitted photon energy of 257 eV, this energy is equal to 512 eV for the curve 2, and 764 eV in the case of the curve 3. These dependences have features of the threshold of generation, as discussed above, and present asymmetric peaks with elongated tails at large values of the field parameter. The probability of the single photon generation decreases sharply with increasing the number of fluxoids that are destroyed in the final state of the superconducting ring.

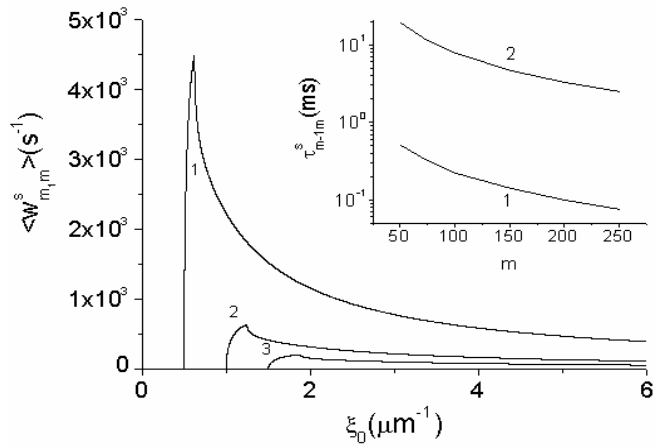

Figure 2. The field parameter dependence of the averaged probability (35) of the supercurrent decay in the channel $m \to m_1$ with the single photon emission from the ring I. The fluxoid number in the initial superconducting state is equal to 101. The curve 1 corresponds to destruction of 1 fluxoid in the final state ($m_1 = m-1$); curve 2 - $m_1 = m-2$ and curve 3 - $m_1 = m-3$. Inset: the fluxoid number dependence of the minimal lifetime of the supercurrent in the decay channel $m \to m-1$. The curve 1 is calculated for the ring I, the curve 2 - for the ring II.

The peaks presented in figure 2, allow to define the maximal probabilities of the single photon generation or, other words, the minimal lifetime of the superconducting currents in the rings. It turned out that the field parameter values, which correspond to positions of the maxima of the peaks, do not depend on the fluxoid number in the initial state. However, the minimum lifetime decreases with increasing the number of fluxoids (curve 1 in the inset to figure 2). With increasing the ring size the probability of the single photon generation is also reduced (curve 2 in the inset to figure 2).

It should be noted that the number of fluxoids that can be trapped in a ring, is limited. The maximal value of this number was obtained in [17] by limiting the supercurrent density at the inner edge of the ring by the depairing current. For the type II superconductors there is also another limitation on this number, which is due to the use of the wave function (3). When the magnetic field created by the supercurrent which is proportional to the fluxoid number, exceeds the first critical field for the type II superconductors, the condensate wave function is more complicated.

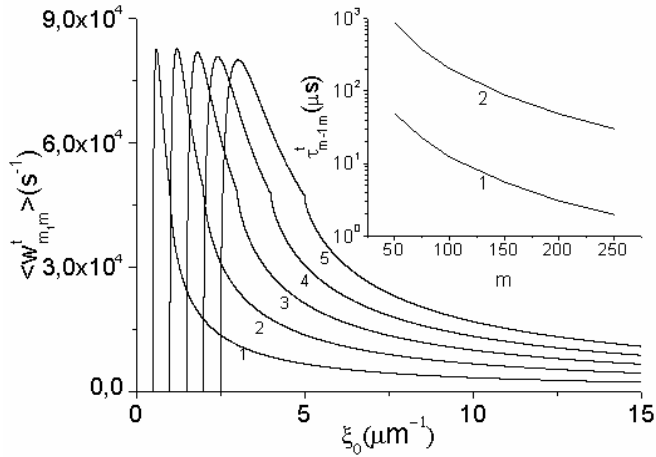

Figure 3. The field parameter dependence of the averaged probability (48) of the two-photon generation by the ring I. The fluxoid number in the initial superconducting state is equal to $m$=101. The curve 1 corresponds to the supercurrent decay channel $m \to m-1$; curve 2 - $m \to m-2$; curve 3 - $m \to m-3$; curve 4 - $m \to m-4$ and curve 5 - $m \to m-5$. Inset: the fluxoid number dependence of the minimal lifetime of the supercurrent in the decay channel $m \to m-1$. The curve 1 is calculated for the ring I, the curve 2 - for the ring II.

The two-photon probability (48) as a function of the field parameter for various supercurrent transitions $m \to m_1$ is presented in figure 3. As in the case of single-photon generation, these dependences have the generation thresholds, and present asymmetric peaks. The peak width increases with the change $m - m_1$ (curves 1-5 in figure 3). A distinctive feature is that the maxima of the two-photon probabilities depend weakly on $m - m_1$. It should be noted that at a given frequency of the coherent field, the field intensity at which the transition $m - m_1 = 5$ becomes allowed, is 25 times greater than the threshold intensity for the transition $m - m_1 = 1$. At the same time the intensity must be kept low to avoid heating the superconducting film irradiated by the coherent electromagnetic wave.

The minimal lifetime of the superconducting current that corresponds to the maximal probability of two-photon emission from the ring, decreases with increasing the number of fluxoids, and increases with increasing the ring size (curves 1 and 2 in the inset to figure 3).

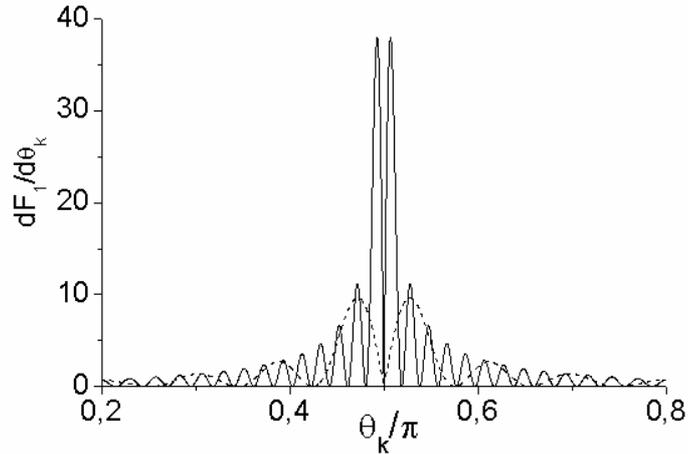

Figure 4. The polar angle spectra of photons emitted at the single-photon supercurrent transition $m \to m-1$. The fluxoid number in the initial superconducting state is equal to $m=101$. The field parameter is $\xi_0 b = 1.2$. The solid curve corresponds to the ring I, dashed curve - the ring II.

Figure 4 shows the polar angle distributions of the photons emitted by the superconducting rings I and II presented in table 1, for the allowed single-photon transition with $m - m_1 = 1$. The distribution is always symmetric with respect to the angle $\theta_k = \pi/2$, at which photon emission is absent (curve 1 in figure 4). Upon increasing both the fluxoid number in the initial state and the change $m - m_1$ oscillations in the polar-angle spectra enhance. However, these oscillations become weaker with the increased ring size (curve 2 in figure 4).

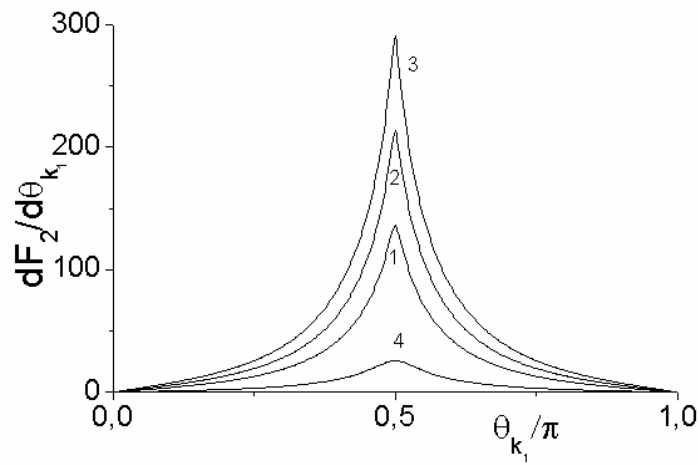

Figure 5. The polar angle spectra of photons emitted at the two-photon generation. For the ring I: curve 1 - $m=101$, $m_1 = m-1$; curve 2 - $m=151$, $m_1 = m-1$; curve 3 - $m=101$, $m_1 = m-2$. For the ring II: curve 4 - $m=101$, $m_1 = m-1$.

For the two-photon generation the polar angle distributions of the emitted photons are presented in figure 5. These distributions determined by (47), are always peaked up at $\pi/2$, and in contrast to the single-photon generation do not contain any oscillations. With increasing the change of the ring energy $E_0(m^2 - m_1^2)$ the height of the peaks increases, but its width reduces slightly (curves 1-4 in figure 5).

It should be noted that, according to the data presented in figures 4 and 5, there are finite probabilities of large track lengths of photons in the material of the superconducting ring. These paths are determined by the difference between the outer and inner radii of the superconducting rings. When the absorption length of photons in the material of the superconducting ring is comparable to or shorter than the average track lengths, the interaction of photons with the superconductor should be taken into account. Of course, the interaction of photons with the superconducting condensate for the considered energies is not significant, and the most probable process of photon scattering is the ionization of core levels of atoms of the superconductor. It can change the yield of photons from rings, but does not lead to an increase of the lifetimes of supercurrent states.

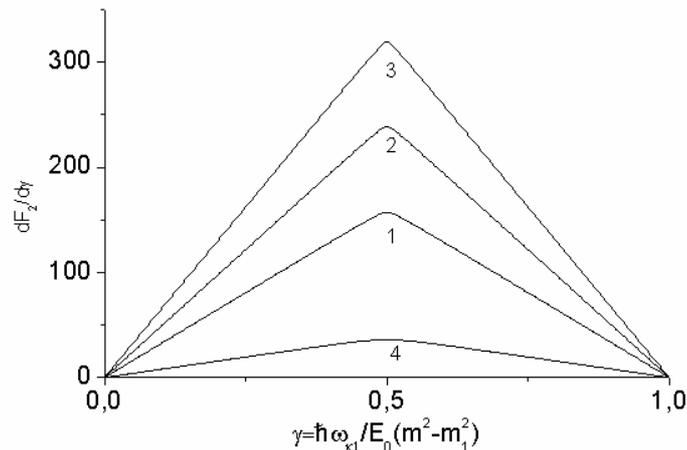

Figure 6. The energy spectra of photons emitted at the two-photon generation. For the ring I: curve 1 - $m=101$, $m_1 = m-1$; curve 2 - $m=151$, $m_1 = m-1$; curve 3 - $m=101$, $m_1 = m-2$. For the ring II: curve 4 - $m=101$, $m_1 = m-1$.

The energy spectra of the photons emitted at the two-photon generation are demonstrated in figure 6. These spectra are determined the function (47) which depends only on the parameter $\beta$ (33). The energy distributions are broad peaks, which are formed by two nearly straight lines (curves 1-4 in figure 6). The maxima of these peaks always fall to the photon energy equal to $E_0(m^2 - m_1^2)/2$. With increasing the change of the ring energy $E_0(m^2 - m_1^2)$ the height of the peaks increases, but its width reduces slightly.

## 8. Conclusion

The microwave field-induced decay of the supercurrent in rings with single-photon and two-photon emission has been studied. This effect can be observed only in the thin-film rings, the thickness of which is less than both the field skin-depth and the London penetration depth. Under this condition the coherent field can cause the collective transitions of all the Cooper pairs involved in the supercurrent.

Comparing the data presented in figure 2 and figure 3, we conclude that the probability of the two-photon generation is much larger than the single-photon one. With decreasing the sizes of the superconductive rings the lifetimes of the allowed transitions decrease, and the ring characteristic energies increase. Despite the fact that the reduction in the ring sizes is accompanied by a decrease in the maximal value of the fluxoid number trapped in the ring, submicron rings appear to be more promising for photon generation.

This effect of the coherent-field-induced generation of photons by the superconducting rings can be used as a new source of photons with energies up to keV range.

**Acknowledgments**